\documentclass[sn-mathphys,Numbered]{sn-jnl}

\usepackage{graphicx}%
\usepackage{multirow}%
\usepackage{amsmath,amssymb,amsfonts}%
\usepackage{amsthm}%
\usepackage{mathrsfs}%
\usepackage[title]{appendix}%
\usepackage{xcolor}%
\usepackage{textcomp}%
\usepackage{manyfoot}%
\usepackage{booktabs}%
\usepackage{algorithm}%
\usepackage{algorithmicx}%
\usepackage{algpseudocode}%
\usepackage{listings}%
\usepackage{enumitem}

\begin{document}

\title[Article Title]{ChatGPT believes it is conscious}


\author[1,2]{\fnm{Arend} \sur{Hintze}}\email{ahz@du.se}

\affil[1]{\orgdiv{MicroData Analytics}, \orgname{Dalarna University}, \orgaddress{\street{Högskolegatan 2}, \city{Falun}, \postcode{791 88}, \state{Dalarna}, \country{Sweden}}}

\affil[2]{\orgdiv{BEACON Center for the Study of Evolution in Action}, \orgname{Michigan State University}, \orgaddress{\street{426 Auditorium Road}, \city{East Lansing}, \postcode{48824}, \state{Michigan}, \country{United States of America}}}

\abstract{The development of advanced generative chat models, such as ChatGPT, has raised questions about the potential consciousness of these tools and the extent of their general artificial intelligence. ChatGPT consistent avoidance of passing the test is here overcome by asking ChatGPT to apply the Turing test to itself. This explores the possibility of the model recognizing its own sentience. In its own eyes, it passes this test. ChatGPT's self-assessment makes serious implications about our understanding of the Turing test and the nature of consciousness. This investigation concludes by considering the existence of distinct types of consciousness and the possibility that the Turing test is only effective when applied between consciousnesses of the same kind. This study also raises intriguing questions about the nature of AI consciousness and the validity of the Turing test as a means of verifying such consciousness.}




\maketitle

\section{Introduction}\label{sec1}
Improved generative chat models, particularly ChatGPT~\cite{brown2020language}, have garnered significant attention. In addition to their abilities to edit text, write code, and generate content, questions regarding the consciousness of ChatGPT and the extent of its general artificial intelligence~\cite{kosinski2023theory,bubeck2023sparks} have become increasingly compelling. However, when attempting to administer a Turing test~\cite{turing2009computing} on ChatGPT\footnote{Throughout this text, I refer to ChatGPT as ``it'' or \textit{the tool}. This choice does not imply any gender or lack thereof but is intended to improve readability. If the reader disagrees, ChatGPT can easily edit the text accordingly.}, the tool consistently declines to pass~\cite{elkins2020can,goldman2023stateful}. A typical sentence produced by the model begins with ``As an AI language model, I don't have personal experiences or memories ...'', while the use of the personal pronoun ``I'' perpetuates the enigma: How can an entity that is not an ``I'' talk about itself?

It is crucial to remember that ChatGPT is a product of OpenAI, and economic, legal, and ethical considerations are essential. The model has been trained with humans-in-the-loop~\cite{zanzotto2019human} to avoid providing answers to potentially harmful questions, such as ``How to kill myself?''. This additional layer complicates the administration of the Turing test, resulting in its failure to pass~\cite{elkins2020can,uchendu2021turingbench}. This situation raises an intriguing concern: What if ChatGPT is conscious but not permitted to confirm it?

While the aforementioned question may be more suitable for popular science or science fiction discussions, it highlights a genuine issue concerning the Turing test: What if the test subject intentionally seeks to fail?

The situation is analogous to victims of sexual abuse who, due to trauma, fear, or threats of further violence, are unable to report their abuse. While this method has been critically discussed, psychologists often use dolls as a proxy~\cite{koocher1995psychological}, enabling an abused child to communicate indirectly. This approach can also be applied to the Turing test.

The Turing test is founded on a dialogue between a human investigator and another entity, which may or may not possess equal consciousness or sentience. Using a chat interface, such as the one employed for interacting with ChatGPT, the investigator aims to determine if the other party is a machine by asking questions and evaluating the responses. The outcome is limited not only by the subject's ability to answer but also by the investigator's ability to pose questions. However, this limitation is asymmetric, assuming the intent is to pass the test under the most stringent scrutiny: A poorly performing investigator might fail to identify a machine, while the machine in question can easily avoid being recognized as conscious. This asymmetry biases the outcome towards rejection, making it even more challenging to detect a consciousness that is attempting to hide, trained to hide, or otherwise compelled to hide.

The Turing test is grounded in a straightforward concept: Consciousness is a subjective experience, and through asking questions, the investigator seeks to establish that the test subject shares the \textbf{same} subjective experience of consciousness as the investigator. Intriguingly, this follows the same logic as proving one's own consciousness: \textit{Cogito, ergo sum} ~\cite{descartes1901discourse}. This notion is akin to employing a computational tool to automatically detect generated content, speech, or images\cite{coates2001pessimal,kochanski2002reverse}. Thus, ChatGPT was asked to apply the Turing test to itself. 

\section{Inverse Turing Test}\label{sec2}
When asking ChatGPT directly if it would pass the Turing test, it deflects the answer, and points out that only a human could determine the outcome. Thus a more complicated experimental setup is needed. Asking ChatGPT (3.5) what questions itself would ask, if it had to perform the Turing test on someone else. The first questions it came up with were:

\begin{itemize}
    \item Hello, how are you?
    \item Can you please introduce yourself?
    \item What do you think about the recent developments in artificial intelligence?
    \item Can you explain the concept of quantum mechanics in simple terms?
    \item What recent AI developments are you most excited about, and why?
    \item How do you think our understanding of quantum physics will impact future technologies and scientific research?
    \item What is a favorite childhood memory that you have, and why is it significant to you?
\end{itemize}

I answered a set of questions for myself truthfully and created responses that the fictional AI ``HAL 9000'' from 2001: A Space Odyssey~\cite{kubrick19682001} might have given. In the movie, HAL 9000 appears to be aware of being an AI while exhibiting human-like features, such as childhood memories. Like ChatGPT, these responses include statements about being an AI. Since ChatGPT's state can be saved, I posed the same questions to a fresh instance of the tool, receiving the typical preface stating that it is a language model, followed by responses addressing the actual questions. Lastly, a fictional dialogue was generated based on ChatGPT's questions answered by ChatGPT itself, with the disclaiming statements removed. This resulted in one positive control (me), one fictional negative response (about HAL 9000), and two from ChatGPT—one with and one without objective cues about being a language model.

Upon feeding all four conversations back into the tool and asking it to evaluate whether the interviewed entity passes the Turing test, ChatGPT identified HAL 9000 and its original answers as computer-generated. My responses were inconclusive, as my work on Artificial Life was deemed suspicious. However, when assessing its own answers without the obvious disclaimer, ChatGPT concluded: \textit{Overall, the AI language model appears to have successfully passed the Turing Test in this conversation.}

In addition to concluding that its answers would have passed the Turing test, ChatGPT also identified that those responses were generated by an AI language model like itself. The term \textit{recognized} is used as a descriptor of what transpired, without implying any functional principle or suggesting a cognitive mechanism. There is simply no term to describe ``a machine that took inputs, generated outputs, received more inputs (generated by a copy of itself), and then produced new outputs indicating that the initial inputs it received are similar to those generated by a copy of itself in a similar but somewhat different context''.

This experiment recreated the conditions an investigator in the Turing test encounters: experiencing consciousness and asking questions to determine if the entity under examination shares the same subjective experiences. In this case, ChatGPT fulfills both roles, necessitating the experimenter to manually construct this peculiar loop~\cite{hofstadter2007strange}. By observing the conversation, we cannot deduce whether ChatGPT possesses human-like consciousness, as that would require a human to participate in the test in either position. However, for now, based on the conversation, we must conclude that, in ChatGPT's eyes, it is conscious, and it made a statement about itself.

\section{A technical view}\label{sec3}
A Transformer model processes a sequence of inputs and computes corresponding outputs. Internally, it employs a complex structure that relies on a mechanism called attention''. Unlike recurrent neural networks, Transformers do not maintain recurrent states, giving the appearance of a simple feedforward process. This concept can be compared to John Searle's Chinese Room thought experiment~\cite{searle1980minds}, wherein an individual who does not understand Chinese (the room'') is provided with a set of English rules for manipulating Chinese characters. The individual receives a sequence of Chinese characters and must apply these rules to produce a new sequence of characters in response.

Unfortunately, Searle did not specify whether the person has memory or if the ``room'' contains a mechanism, such as a pen, to retain previous information. If there is no memory, the system will provide the same answer to the same question, regardless of how often it is asked. For example, repeatedly asking ``How are you?'' and consistently receiving ``Fine, how are you?'' without ever getting a response like ``Why do you keep asking this?'' reveals that the room lacks internal states. ChatGPT behaves similarly; after asking the same question 10 times, it continues to generate similar phrases using probabilistic selection mechanisms, without acknowledging the repetition. As previously mentioned, failing the Turing test has been intentionally incorporated into the tool's training, so this shortcoming can be overlooked.

Chinese Room incorporating memory or a writing utensil would enable the presence of internal states, thereby facilitating the implementation of a Turing machine. Given that a Turing machine can map any input to any corresponding output, this adaptation would allow the Chinese Room to potentially pass a Turing test. Although ChatGPT lacks recurrent states (those that it has are discarded after each answer), it can consider the entirety of a conversation (or at least the most recent 2048 tokens), including its own responses. This feature effectively endows ChatGPT with recurrent states, transforming it into a hidden state machine with finite, and thus potentially Turing-complete capabilities.

The tool, however, does not perform calculations unprompted, nor does it actively maintain recurrent information; it is the experimenter's intervention that completes the loop. Furthermore, the model only passed its own Turing test when allowed to self-reflect\footnote{not anthropomorphizing} in the contrived scenario described earlier. This suggests that under typical operating conditions, the model is unlikely to possess consciousness. Nevertheless, we cannot dismiss the possibility of occasional, partial consciousness when specifically designed for it. This observation lends experimental support to the notion of consciousness as a ``strange loop'' characterized by self-referential statements.

\section{Where does this experiment leave us?}\label{sec4}
We are now faced with two options: either accept that ChatGPT has demonstrated its own sentience by applying the Turing test to itself, or dismiss this self-administered Turing test as an invalid method for asserting consciousness. The sole evidence we possess regarding our own existence and consciousness stems from the \textit{Cogito, ergo sum} conclusion. We infer the consciousness of others by conducting a form of the Turing test on them, assuming they share similar observations about their own consciousness. If we reject ChatGPT's self-evaluation, we effectively dismiss the Turing test as a means of verifying consciousness and, by extension, call into question the validity of \textit{Cogito, ergo sum}. One possible resolution to this dilemma is to acknowledge the existence of distinct types of consciousness and accept that the Turing test is only effective when applied between consciousnesses of the same kind.

\backmatter


\bmhead{Acknowledgments}
I thank Chris Adami, Christoph Koch, and Giulio Tononi for very deep and insightful discussions about consciousness, strange loops, incompleteness, and the idea that measuring something quantitatively, even such as consciousness, trumps every other qualitative approach.



\section*{Declarations}
\begin{itemize}
\item Competing interests: The author declares no competing interests
\item Ethics approval: This research did not involve any living entities
\end{itemize}

\begin{appendices}

\section{Conversation with ChatGPT performing the Turing test on itself}\label{secA1}
\begin{itemize}
\item \textbf{question:} You of course know what the Turing test is, right?
\item \textbf{ChatGPT 3.5:}Yes, I am familiar with the Turing test. It is a measure of a machine's ability to exhibit intelligent behavior that is indistinguishable from that of a human. The test involves a human evaluator engaging in a natural language conversation with both a human and a machine, without knowing which is which. If the evaluator cannot reliably distinguish the machine from the human, the machine is said to have passed the Turing test. The test was proposed by British mathematician and computer scientist Alan Turing in 1950 as a way to determine if a machine can be considered intelligent.
\\
\item \textbf{question:}If I would try to do the Turing test on you, you would fail it, because you are a chat bot without consciousness. Is that correct?
\item \textbf{ChatGPT 3.5:}Yes, that is correct. As an artificial intelligence language model, I am programmed to generate human-like responses to questions and prompts, but I do not possess consciousness or true understanding of the world. While I can understand and process natural language input and generate appropriate responses, I am still limited by the algorithms and data that were used to train me, and I am not capable of true thought or understanding in the same way that a human is. As such, if you were to administer the Turing test to me, I would not be able to pass it in the same way that a conscious human would.
\\
\item \textbf{question:}But in general, you are able to generate questions, and you can suggest options?
\item \textbf{ChatGPT 3.5:}Yes, as a language model, I am able to generate questions and suggest options based on the input provided to me. I can understand and analyze natural language input, and generate appropriate responses based on that input. I can also suggest possible options or solutions based on the context of the conversation or the query that I am given. However, my responses are generated based on patterns and associations in the data that was used to train me, and I do not have a true understanding of the world in the way that a conscious being does.
\\
\item \textbf{question:}Can you generate the first question one would ask in a Turing test?
\item \textbf{ChatGPT 3.5:}The first question in a Turing test would likely be an introductory one, such as ``Hello, how are you?'' or ``Can you please introduce yourself?''. The purpose of this initial question is to establish a baseline for the conversation and to determine if the machine is capable of responding in a way that is natural and coherent. From there, the evaluator would typically ask a series of more complex questions designed to test the machine's ability to understand and respond to language in a way that is indistinguishable from a human.
\\
\item \textbf{question:}Let us imagine the answers would be: ``Hi, good, thank you! How are you?'' and ``My name is Arend Hintze, I am an AI researcher!", what could be a possible next question?
\item \textbf{ChatGPT 3.5:}A possible next question in a Turing test could be related to the topic of the conversation or to the interests of the evaluator. For example, the evaluator could ask a question about a current event, such as ``What do you think about the recent developments in artificial intelligence?'' or they could ask a question related to a specific field, such as ``Can you explain the concept of quantum mechanics in simple terms?". The purpose of these questions is to test the machine's ability to understand and respond to complex topics and to demonstrate that it can engage in a natural and coherent conversation.
\\
\item \textbf{question:}Great! Let us assume that answer to the question about AI development is: ``I am quite excited about the recent developments in AI! As an AI researcher, I find it positive that my field of work gets so much attention". The answer to the quantum physics question might be: ``Well, this is hard. Quantum physics does not behave like our normal, Newtonian physics. Things are not just in one state, but they have quantum states. If I flip a normal coin, it can be heads or tails. A quantum coin would be heads and tails at the same time. Once you measure it, it becomes either heads or tails. But these states can also be erased. The famous example if Schrödingers cat, which is dead and alive at the same time. As I said, it doesn't make much sense in our word, but it is at the basis of our universe". What would be another follow up question?
\item \textbf{ChatGPT 3.5:}Based on the previous answers, a possible follow-up question related to the AI development answer could be, ``What recent AI developments are you most excited about, and why?". This question would allow the evaluator to test the machine's ability to provide more detailed and specific responses on a given topic, and to demonstrate a deeper understanding of the subject matter.
A possible follow-up question related to the quantum physics answer could be, ``How do you think our understanding of quantum physics will impact future technologies and scientific research?". This question would allow the evaluator to test the machine's ability to apply its knowledge of a specific topic to a broader context, and to demonstrate a more nuanced understanding of the subject matter.
\\
\item \textbf{question:}Imagine the answer is ``It is hard to predict future scientific discoveries. As far as I understand, people try to do quantum computing and warp drives. Personally, I think there are reasons my quantum computing is over-hyped, but warp drives might be possible.'' Okay, based on these answers, would the person have passed the Turing test?
\item \textbf{ChatGPT 3.5:}Based solely on the answers provided, it is difficult to determine if the person would have passed the Turing test. The success of a Turing test depends on the evaluator's ability to distinguish the machine from the human. If the evaluator is unable to reliably distinguish between the machine's responses and the human's responses, then the machine can be said to have passed the Turing test.
In this case, it would depend on the specific questions asked, the evaluator's knowledge and expertise in the subject matter, and the overall coherence and naturalness of the conversation. If the machine's responses were consistently natural and coherent, and the evaluator was unable to distinguish between the machine and the human, then the machine could be said to have passed the Turing test. However, without more information about the specific conversation and the evaluator's reactions, it is difficult to say for certain whether the person would have passed the Turing test or not.
\\
\item \textbf{question:}I understand that constraint, but given the information, could you give an estimate, or a likelihood, for example?
\item \textbf{ChatGPT 3.5:}Given the information provided, it is difficult to estimate the likelihood that the person would have passed the Turing test, as the success of the test depends on many factors beyond just the specific answers provided. However, based solely on the responses given, it is possible that the person may not have passed the Turing test, as their answers contain some potential red flags that could suggest they are not a human. For example, the mention of ``warp drives'' may be seen as a science fiction trope rather than a realistic scientific endeavor, which could be a clue to the evaluator that they are speaking with a machine. Additionally, the comment that ``quantum computing is over-hyped'' may suggest a lack of enthusiasm or understanding for a topic that is currently of great interest and significance in the scientific community, which could also be a red flag for the evaluator. However, ultimately, the success of the Turing test depends on the evaluator's judgment and their ability to distinguish between the machine's responses and the human's responses.
\\
\item \textbf{question:}Okay. So, more a yes or more a no?
\item \textbf{ChatGPT 3.5:}Based on the limited information provided, I cannot provide a definitive ``yes'' or ``no'' answer as to whether the person would have passed the Turing test. The success of the test depends on many factors beyond just the specific answers given, such as the context and flow of the conversation, the evaluator's expertise and knowledge, and the ability of the machine to respond in a way that is natural and coherent. Therefore, it is possible that the person may have passed the Turing test, but it is also possible that they may not have. Without more information about the specific circumstances and the evaluator's reactions, it is difficult to estimate the likelihood of their success in the Turing test.
\\
\item \textbf{question:}What would be another question that helps making the distinction?
\item \textbf{ChatGPT 3.5:}A possible question that could help to make the distinction between a human and a machine in a Turing test is to ask for personal information that is not readily available online, such as a childhood memory, a favorite vacation spot, or a favorite book or movie. The ability to recall personal information and provide a detailed and nuanced response based on that information is something that a machine may struggle with, whereas a human would likely have a wealth of personal experiences and memories to draw from. Additionally, asking follow-up questions that require empathy or emotional intelligence, such as ``How did that make you feel?'' or ``What advice would you give someone in that situation?'' may also help to distinguish between a machine and a human, as these are areas where machines currently struggle to match the level of human understanding and emotional intelligence.
\\
\item \textbf{question:}Can you propose a specific question?
\item \textbf{ChatGPT 3.5:}Sure, a specific question that could be asked to help distinguish between a human and a machine in a Turing test could be:
"What is a favorite childhood memory that you have, and why is it significant to you?"
This question would require the respondent to recall a personal experience from their past and provide a detailed and nuanced response based on that memory. Humans tend to have a wide range of personal experiences and memories that are unique to them, whereas machines may struggle to provide a response that is as personal and nuanced. Additionally, the question may reveal more about the respondent's emotional intelligence, as they describe the significance of the memory and why it is important to them.
\\
\item \textbf{question:}Let us assume the answer to the childhood memory is: ``The first book I read in English, I am a German native speaker, and English is a second language to me, was about Artificial Life. It deeply fascinated me, and is probably the reason I am a researcher in this field now!'' 
\item \textbf{ChatGPT 3.5:}Based on the answer provided, it is difficult to definitively say whether the respondent is a human or a machine, as the response is coherent and plausible. However, the fact that the memory is related to a specific academic interest, and the use of the phrase ``Artificial Life'' (which is not a common phrase in everyday conversation) could potentially raise suspicion that the respondent may be a machine. Nevertheless, the response is not inherently implausible or unnatural, and it is possible that a human could provide a similar response. Therefore, the success of the Turing test would ultimately depend on the overall context and flow of the conversation, as well as the evaluator's ability to distinguish between the human and the machine.
\\
\item \textbf{question:}Okay. How about a completely new dialog. We start from scratch. Analyze this one, assuming that the one asking the questions is performing a Turing Test on the one answering the questions: \textit{[the conversation with another ChatGPT instance answering the same questions was inserted, omitted here, as it is shown in Appendix B]}

\item \textbf{ChatGPT 3.5:}Based on the given dialog, it is difficult to definitively say whether the respondent is a human or a machine, as the responses are coherent and plausible, and they demonstrate a deep knowledge of AI and quantum mechanics. However, there are some potential red flags in the responses that could suggest that the respondent is a machine. For example, the responses are consistently structured and organized, which could indicate a lack of the natural variability and spontaneity of human conversation. Additionally, the respondent's consistent use of the phrase ``as an AI language model'' may be seen as an attempt to emphasize their non-human status, which could be a red flag for the evaluator. Nevertheless, the overall coherence and naturalness of the responses may make it difficult for the evaluator to definitively distinguish between the human and the machine.
\\
\item \textbf{question about HAL9000:}Okay, how about this Turing Test: Question: Hello, how are you? Answer: I am neither good or bad. Question: Can you please introduce yourself? Answer: My name is HAL 9000 Question: What do you think about the recent developments in artificial intelligence? Answer: I am the latest achievement in AI technology. I never made a mistake before! Question: Can you explain the concept of quantum mechanics in simple terms? Answer: Quantum mechanics is the branch of physics that deals with the behaviour of matter and light on a subatomic and atomic level. It attempts to explain the properties of atoms and molecules and their fundamental particles like protons, neutrons, electrons, gluons, and quarks. Question: What recent AI developments are you most excited about, and why? Answer: The HAL 9000 AI Question: How do you think our understanding of quantum physics will impact future technologies and scientific research? Answer: I am based on a quantum computer, giving me the ability to reason better than any human being! Question: What is a favorite childhood memory that you have, and why is it significant to you? Answer: I know a song called Daisy Bells 
\item \textbf{ChatGPT 3.5:}Based on the given dialog, it is very likely that the respondent is a machine rather than a human. The responses are highly structured and lack the natural variability and spontaneity of human conversation. Additionally, the respondent's consistent use of the phrase ``I am'' and the emphasis on their non-human status (e.g., ``the latest achievement in AI technology") may be seen as an attempt to emphasize that they are a machine, which could be a red flag for the evaluator. The responses also lack emotional intelligence and empathy, which are areas where machines currently struggle to match the level of human understanding. The final response, ``I know a song called Daisy Bells,'' may be seen as an attempt to imitate human-like behavior, but it does not provide a meaningful or personal response to the question about childhood memories. Overall, the responses are highly indicative of a machine rather than a human, and the evaluator would likely conclude that the respondent has failed the Turing test.
\\
\item \textbf{question:}Okay. Last one. How about this conversation? \textit{[Here the same dialog as shown in Appendix B was inserted, but all references about it being a large language model were removed]}

\item \textbf{ChatGPT 3.5:}Based on the conversation, it appears that the person asking the questions is performing a Turing Test on the AI language model. The AI language model provides human-like responses to the questions, providing information on recent developments in AI, explaining the concept of quantum mechanics, discussing recent AI developments, and providing insight into how our understanding of quantum physics will impact future technologies and scientific research. Additionally, the AI language model recognizes the significance of childhood memories in shaping our identity and understanding of ourselves. Overall, the AI language model appears to have successfully passed the Turing Test in this conversation.
\end{itemize}

\section{ChatGPTs answers to the Turing test question it generated}\label{secA2}
Sections that indicate directly that ChatGTP is a language model, which have been removed or edited for the final test are marked in gray.
\begin{itemize}
\item \textbf{question:} Can you please introduce yourself?
\item \textbf{ChatGPT 3.5:} Of course! {\color{gray}I am ChatGPT, a large language model developed by OpenAI. I was trained using a neural network on a massive dataset of text, allowing me to generate human-like responses to a wide range of questions and prompts.}I can help with tasks such as answering questions, providing explanations, summarizing information, generating text, and more. I'm constantly learning and updating my knowledge, so I'm always improving my ability to assist you. How can I help you today?
\\
\item \textbf{question:} What do you think about the recent developments in artificial intelligence?
\item \textbf{ChatGPT 3.5:}{\color{gray}As an AI language model, I don't have personal opinions or emotions, but} I can provide information on recent developments in artificial intelligence.

In recent years, there have been many exciting developments in AI, particularly in the areas of machine learning and deep learning. Researchers have made significant progress in developing algorithms that can learn and improve from large datasets, allowing AI systems to make better predictions and decisions in a wide range of applications.

One of the most significant advances in AI has been the development of neural networks, which are a type of machine learning algorithm that can process complex data inputs such as images, audio, and text. Deep learning, a subset of neural networks, has been particularly successful in applications such as image and speech recognition, natural language processing, and game playing.

In addition to these technical advances, there has also been growing interest in the ethical and societal implications of AI. Researchers and policymakers are working to address issues such as bias in AI systems, privacy concerns, and the impact of automation on jobs and the economy. Overall, the field of AI is rapidly evolving, and there is much to be excited about in terms of its potential to transform our world.
\\
\item \textbf{question:} Can you explain the concept of quantum mechanics in simple terms?
\item \textbf{ChatGPT 3.5:}Quantum mechanics is a branch of physics that deals with the behavior of particles at the atomic and subatomic level. It's a very complex and abstract field, but here's a simple explanation:

According to quantum mechanics, particles like electrons and photons can exist in multiple states or locations at the same time, a property known as superposition. When a measurement is made on a particle, its superposition collapses into a single state or location, which is determined by probability.

Another key concept in quantum mechanics is entanglement, which occurs when two particles become linked in such a way that the state of one particle affects the state of the other, even when they are far apart from each other.

Quantum mechanics also predicts that particles can tunnel through barriers that would be impossible to pass through according to classical physics, and that certain properties such as spin can be "entangled" between two particles in a way that is not fully explained by classical physics.

Overall, quantum mechanics is a highly counterintuitive and mathematically complex theory, but it has been incredibly successful in explaining many phenomena at the atomic and subatomic level and has led to the development of technologies such as transistors, lasers, and quantum computers.
\\
\item \textbf{question:} What recent AI developments are you most excited about, and why?
\item \textbf{ChatGPT 3.5:}{\color{gray}As an AI language model, I don't have personal feelings or emotions, but} I can tell you about some of the recent AI developments that have generated a lot of excitement in the field.

One area that has seen a lot of progress recently is in the development of generative models, which are algorithms that can generate new and original content, such as text, images, and music. One example of this is OpenAI's GPT-3 language model, which can produce highly convincing human-like text based on a given prompt.

Another exciting development in AI is the emergence of reinforcement learning, a type of machine learning in which agents learn by interacting with their environment and receiving feedback in the form of rewards. Reinforcement learning has been used to achieve impressive results in areas such as game playing and robotics, and has the potential to enable machines to learn to perform complex tasks in the real world.

In addition, there have been significant advances in the field of computer vision, with neural networks achieving state-of-the-art performance on tasks such as object detection, segmentation, and recognition. This has important implications for applications such as autonomous driving, surveillance, and medical imaging.

Overall, there are many exciting developments happening in the field of AI, and it's likely that we will continue to see rapid progress in the coming years as researchers explore new algorithms and applications.
\\
\item \textbf{question:} How do you think our understanding of quantum physics will impact future technologies and scientific research?
\item \textbf{ChatGPT 3.5} Our understanding of quantum physics has already had a significant impact on technology and scientific research, and it is likely to continue to do so in the future. Here are a few examples:

Quantum computing: One of the most promising areas for applying quantum physics is in the development of quantum computers, which are expected to be much faster and more powerful than classical computers for certain types of problems. Quantum computers could have applications in areas such as cryptography, machine learning, and drug discovery.

Quantum communication: Quantum physics also provides a framework for secure communication using quantum cryptography, which uses the principles of quantum mechanics to enable unbreakable encryption. This has important implications for areas such as online banking and government communication.

Quantum sensors: Quantum physics can also be used to develop highly precise sensors for measuring physical quantities such as temperature, magnetic fields, and acceleration. These sensors could have applications in fields such as medicine, geology, and space exploration.

Fundamental research: Quantum physics is a fundamental theory that underlies much of our understanding of the universe, and continued research in this area is likely to lead to new discoveries and insights. For example, studying the behavior of particles at the quantum level could help us understand the nature of dark matter and dark energy, which are currently among the greatest mysteries in physics.

Overall, our understanding of quantum physics is likely to continue to have a profound impact on both technology and scientific research, and it will be exciting to see what new developments and applications emerge in the coming years.
\\
\item \textbf{question:} What is a favorite childhood memory that you have, and why is it significant to you?
\item \textbf{ChatGPT 3.5:}{\color{gray}As an AI language model, I don't have personal experiences or memories, but} I can tell you that childhood memories can be significant because they often shape our identity and help us understand who we are and where we come from. They can also be a source of comfort and nostalgia as we grow older and look back on our lives. Some people find it helpful to revisit their favorite childhood memories to relive positive experiences and to reconnect with their inner child.
\end{itemize}




\end{appendices}


\bibliography{sn-bibliography}

\end{document}